\documentclass[journal,comsoc]{IEEEtran}

\usepackage[T1]{fontenc}

\ifCLASSINFOpdf
   \usepackage[pdftex]{graphicx}

  \DeclareGraphicsExtensions{.pdf,.jpeg,.png}
\else

\fi

\usepackage{amsmath}

\usepackage[cmintegrals]{newtxmath}

\usepackage[caption=false]{subfig}
\usepackage{algorithm}
\usepackage[noend]{algpseudocode}

\usepackage{graphicx}
\usepackage{latexsym}
\usepackage{amsmath,epsfig}
\usepackage{epstopdf}
\usepackage{caption}
\usepackage{multirow}
\usepackage{booktabs,siunitx}
\usepackage{lscape}
\usepackage{url}
\usepackage{graphicx}
\usepackage{epsfig}
\usepackage{graphicx}
\usepackage{latexsym}
\usepackage{amsmath}
\usepackage{amssymb}
\usepackage{mhsetup}
\usepackage{mathtools}
\usepackage{mathrsfs}
\usepackage{float}
\usepackage{xfrac}
\usepackage{epsfig}
\usepackage[english]{babel}
\usepackage{subfig}

\begin{document}

\title{Time Synchronization in Wireless Sensor Networks based on Newton's Adaptive Algorithm}

\author{Ramadan~Abdul-Rashid and Azzedine~Zerguine \\
	
Electrical Engineering Department, King Fahd University of Petroleum and Minerals, Dhahran, Saudi Arabia \\
Email: {ram.rashid.rr@gmail.com, azzedine@kfupm.edu.sa}
}



\maketitle

\begin{abstract}
This paper proposes a  novel time synchronization protocol inspired by the adaptive Newton search algorithm. The clock model of nodes are modeled as an adaptive filter and a pairwise steady state and convergence analyses are presented. A protocol is derived from the derived algorithm and compared experimentally to Gradient Descent Synchronization(GraDeS) protocol and the Average Proportional Integral Synchronization (AvgPISync) protocol. Experimental results showed similar error performance as AvgPISync but far outperformed both protocols in terms of convergence time. The protocol is lightweight, robust and simple to implement. 
\end{abstract}


%
\IEEEpeerreviewmaketitle

\section{Introduction}
Wireless Sensor Networks (WSNs) as distributed systems used for several precision and sensitive sensing and instrumentation applications require network nodes to be as closely synchronized in time as possible. Several factors ~\cite{jeremy_elson_wireless_2003}, like the tight link between sensors and the physical world, the scarcity of energy for deployed nodes, the need for large scale deployment, decentralized topologies and unpredictable and intermittent connectivity between network nodes necessitate an accurate, flexible and robust time synchronization for wireless sensor networks. Although most traditional networks depend on physical time for time synchronization, WSN applications such as object tracking, consistent state updates, and distributed beamforming make it impractical to depend on logical physical time for WSN time synchronization. A myriad of studies has gone into developing different methods of time synchronization of WSNs with the goal of optimizing networks parameters like energy utilization, precision, lifetime, scope, scalability, size and cost.

Several distributed algorithms  \cite{schenato_average_2011} have been suggested to solve the synchronization problem. Several of the methods adopted for time synchronization in wireless sensor networks normally assume specific stationary network connectivity \cite{maroti_flooding_2004}, synchronous update \cite{schenato_average_2011}, a multi-hop communication \cite{yildirim2016gradient} and/or a node labeling scheme ~\cite{ he_time_2014-1}, etc. Hence time synchronization protocols operating under such assumptions will not effectively synchronize nodes under these stated conditions and would have to constantly resynchronize to ensure global network synchronization which sharply depletes node power. 

In the Proportional Integral Synchronization (PISync) \cite{yildirim_pi} and Gradient Descent Synchronization (GraDes) protocols \cite{yildirim2016gradient}, a more adaptive approach to the synchronization problem is employed. Whereas in PISync the problem is modeled in a control framework, where PI controller is used to update the relevant clock parameters so as to achieve synchronization, in \cite{yildirim2016gradient}, the synchronization process is formulated as an optimization problem and the gradient descent algorithm is used to adapt the clock speed in each synchronization round. This method show a more nuance approach, in term of computation overhead, convergence time and protocol robustness as compared to the traditional distributed and centralized protocols. The methodology used to device this protocol inspires our proposed protocol where we sort to further improve the synchronization convergence time and accuracy by adopting the Newton Adaptive Search algorithm. The Newton algorithm differs from the gradient descent in that, the gradient descent algorithm ,  it is based on a second-order approximation of error surface as compared to the first order approach used in the gradient descent algorithm. Although the gradient descent algorithm is simple in design and implementation, the Newton algorithm far outperforms it in terms of convergence time.

The key contributions of this paper are outlined below.
\begin{enumerate}
	\item Proposal of a novel method for time synchronization inspired by the Newton Adaptive Algorithm.
	\item Pairwise steady-state and convergence analysis of the proposed algorithm
	\item Present a practical working protocol based on the proposed algorithm
	\item Experimental comparative evaluation on the performance of our suggested synchronizer against the Gradient Descent Time Synchronization Protocol in terms local and global synchronization errors and convergence time.
\end{enumerate}

The rest of the paper is organized as follows. Section III discusses the system model Section IV presents a mathematical analysis of of the proposed method of synchronization and section V presents the suggested method for it practical realization. Practical experimental comparative evaluation of the synchronization protocols are then presented in Section VI. The paper concludes in Section VII.

\section{System Model}
\begin{figure}[ht] 
	\centering    
	\includegraphics[width=0.450\textwidth]{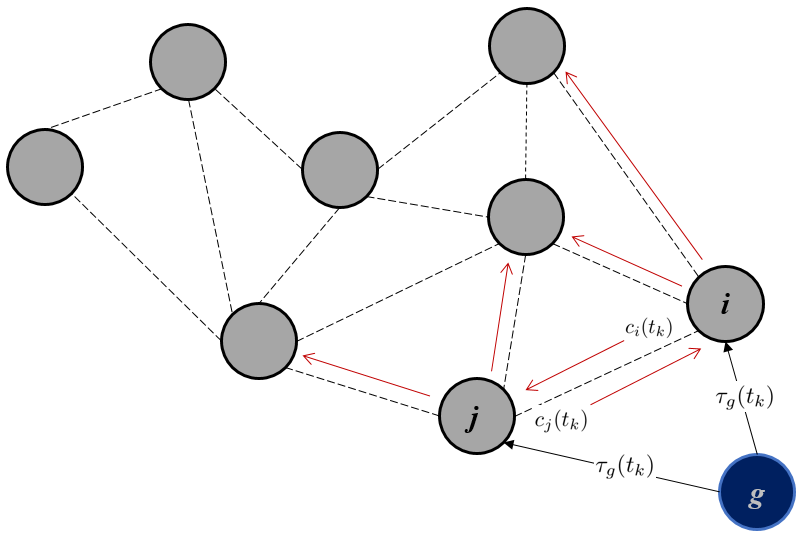}
	\caption[Model Network for Clock Synchronization]{Model Network for Clock Synchronization: where through neighborhood clock information exchange between nodes is used to synchronize each nodes clock to the clock of a gateway node, $\tau_g$.}
	\label{fig:net}
\end{figure}

In a sparsely distributed Wireless Sensor Network (WSN), ordinary nodes spread across a relatively localized area or sensor field periodically report their measurements' or data to an anchor or gateway node. Unlike the ordinary nodes with limited energy, bandwidth and memory, most gateway nodes have a bulk of these resources. Additionally in many wireless sensor network architectures, the gateway node has access to a stable and precise clock reference, for example a GPS receiver \cite{sommer_gradient_2009,akl_investigation_2011}. In the absence of an external clock source, the hardware clock of a this node serve as a source, whose clocks' is tracked by all other sensor nodes for synchronization as shown in Figure \ref{fig:net}.

\subsection{Network Model}
Assume a WSN has symmetric links and hence can be represented by an undirected graph $\mathcal{G}=(\mathcal{V},\mathcal{E})$. In this model, we represent the sensor nodes of the network as the vertex set, $\mathcal{V}=\{v_i|i = 1,2,\hdots ,N\}$, where $N = |\mathcal{V}|$ is the cardinality of $\mathcal{V}$ and the working network connectivity between these nodes as an edge set, $\mathcal{E} \ \text{such that} (v_i,v_j) \in \mathcal{E}$ if nodes $i$ and $j$ can send information to each other. Such nodes that directly communicate with node $i$ are referred to as the neighborhood nodes of $i$ and represented by the set, $\mathcal{N}_i = \{v_j|\mathcal{E}_{i,j}\in \mathcal{E}\} $.

\subsection{Clock Model}
In a WSN, each node is equipped with a hardware clock, $\tau$ which is fashioned using a crystal oscillator. Since the nominal operating oscillators of these oscillators are subject to changes due to changes in temperature, and aging, the hardware clocks exhibit drifts. We adopt the definition of a hardware clock given in \cite{yildirim_external_2014,yildirim_pisync}, where the hardware clock, $\tau_i$ of an arbitrary WSN node $i$ at time $t>t_0$ is modeled in terms of an initial clock value, $\tau_i(t_0)$ and the oscillator frequency, $f(\zeta) \in [\hat{f}-f_{max},\hat{f}+f_{max}]$ as

\begin{equation}\label{eq1}
\tau_i(t) = \tau_i(t_0) + \int_{t_0}^{t} f(\zeta)d\zeta
\end{equation}

where, $f_{min}$ and $f_{max}$ are respectively, the lower and upper bounds of the nominal frequency of the oscillator, $\hat{f}$ and the drift dynamics of the clock is modeled as 

\begin{equation}
f(t) = \hat{f} + r(t)
\end{equation}

where $r(t)$ is uniformly distributed between $-f_{max}$ and $+f_{max}$.

Since the clock parameters of a hardware clock cannot be adjusted manually, each node also maintains a logical clock whose value is a function of the current hardware clock value, $\tau_i(t)$ and a logical clock rate, $\Delta(t)$. This logical clock $c(t)$ of is represented as

\begin{equation}
c(t) = c(t_0) + \Delta (t)[\tau (t) - \tau (t_0)]
\end{equation}

To specify this update rule based on the network model in Figure \ref{fig:net}, at a communication instant,$k$ and latest update time, $t_k$ a node, $i \in \mathcal{V}$ updates its next logical clock as

\begin{equation}\label{eq4}
c_i(t_k^{+}) = c_i(t_k) + \Delta_i(t_k^{+})[\tau_i(t_k^{+}) - \tau_i (t_k)]
\end{equation} 

\begin{figure}[ht] 
	\centering    
	\includegraphics[width=0.40\textwidth]{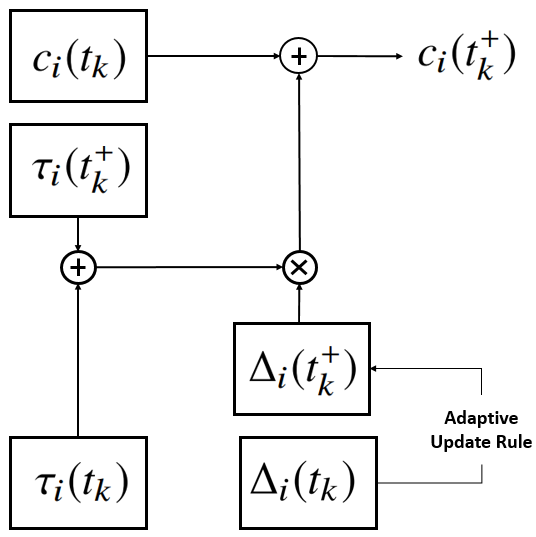}
	\caption[Suggested Clock Setup for Node Synchronization]{Suggested Clock Setup for Node Synchronization}
	\label{fig:model}
\end{figure} 

The logical clock rate, $\Delta_i(t_k) = \frac{d}{dt_k} [c_i(t_k)]$ is an estimation of the relative frequency $\frac{\hat{f}}{f_i(t)}$ and its current value indicates whether the the logical clock is slowed or quickened. Using this evolving model, the clocks of nodes in a WSN can be synchronized by using the current updates of the offset value, $c_i(t_k)$ and the clock rate $\Delta_i(t_k)$ to iteratively maintain the logical clock of node $i$. To obtain a robust synchronization model, an adaptive algorithm is adopted for the update of $\Delta_i$ as shown in Figure \ref{fig:model}.  We can rewrite (\ref{eq4}) as:

\begin{equation}\label{eq5}
\bar{c}_i(t_k) = \Delta_i(t_k^{+}) \bar{\tau}_i(t_k)
\end{equation}

where $\bar{c}_i(t_k) = c_i(t_k^{+}) - c_i(t_k)$ and $\bar{\tau}_i(t_k) = \tau_i(t_k^{+}) - \tau_i (t_k)$.

%
Assuming $M < N = |\mathcal{V}|$ nodes update their clocks at $t_k$, with a little abuse of notation, say $k = t_k$, a generalized form of (\ref{eq5}) can be written as,
\begin{equation}\label{eq6}
C(k) = \bar{\Delta}(k) T(k)
\end{equation}

where $T(k) = [\bar{\tau}_1(k), \hdots, \bar{\tau}_M(k) ]$, $C(k) = [\bar{c}_1(k), \hdots , \bar{c}_M(k)]$ and $\bar{\Delta}(k) = [\Delta_1(k), \hdots , \Delta_M(k)]$

\section{Proposed Method of Synchronization}
In our suggested synchronization framework, the synchronization problem is thought of as an estimation problem, where at each instant, $k$, all nodes try to track and estimate the clock of the gateway clock, $\tau_g(k)$ which is assumed to be perfectly ticking, i.e.
\begin{equation}
\tau_g(k) = k \times B
\end{equation}
where B is the communication rate of the nodes

\begin{figure}[ht] 
	\centering    
	\includegraphics[width=0.40\textwidth]{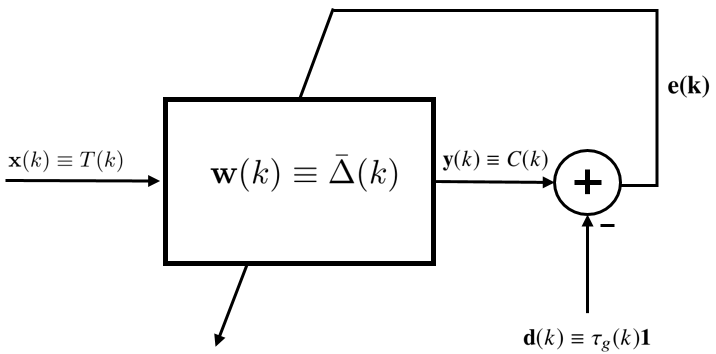}
	\caption[Model Framework for Clock Synchronization]{Model Framework for Clock Synchronization}
	\label{fig:absmodel}
\end{figure}

This system can be abstracted into a linear adaptive update model where the a training set $\{\textbf{x}(k),d(k)\} \equiv \{T(k),\hat{\tau}_g(k)\}$ is used for the update and global synchronization of the WSN. This abstraction is illustrated by Figure \ref{fig:absmodel}. In this model, we utilize the error, $e(k)$ between each node's logical clock with respect to the gateway clock for the adaptive update of clock parameters as given in  (\ref{eq7}).

\begin{equation} \label{eq7}
e(k) =  \bar{c}_i(k) - \tau_g(k) 
\end{equation}

A clear equivalence is therefore observed between the adaptive filter in Figure \ref{fig:absmodel} and the synchronization problem. This equivalence is illustrated by \ref{model};
\begin{equation}\label{model}
\textbf{d}(k) \equiv \tau_g(k)\textbf{1}; \ \textbf{w}(k) \equiv \bar{\Delta}(k);  \ \textbf{x}(k) \equiv T(k); \ \textbf{y}(k) \equiv C(k) \
\end{equation}
where $\textbf{1} = [1,1, \hdots ,1]^{T}$

This abstraction is important because, it allows for the use of any stochastic gradient algorithm for clock rate adaptation. 

\subsection{Pairwise Synchronization to Gateway Node}

Applying Newton's for the update of the logical clock rate of node $i$, we have

\begin{equation} \label{eq9}
\Delta_i (t_k^{+}) = \Delta_i (t_k) - \mu [H(t_k)]^{-1}g(t_k)
\end{equation}

where $g(t_k) = \frac{\partial J(\Delta_i)}{\partial \Delta_i}$ and $H(t_k) = \frac{\partial g(\Delta_i)}{\partial \Delta_i}$

The cost function $J$ that is to minimized is taken as the square error, defined as: 

\begin{equation}
J(t_k) = e^{2}(t_k)
\end{equation}

Assuming node $i$ is a is in communication range of the gateway node, $g$, i.e. $\{g |\mathcal{E}_{i,g} \in \mathcal{E}\}$ and there is a transmission delay of $\beta$ between $i$ and $g$, then at time $t_k$, node $i$ receives $\tau_g = kB + \beta_k$ and hence  (\ref{eq7}) becomes

\begin{equation} \label{eq8}
e(t_k) =  c_i(t_k) - kB - \beta_k = \Delta_i(t_k^{+})\bar{\tau}_i(t_k^{+}) - kB - \beta_k
\end{equation}

In \cite{He_2014}, using the central limit theorem, $\beta_k$ is modeled as a zero-mean Gaussian distributed random variable with variance, $\sigma^2_{\beta}$.

Using  (\ref{eq4}), we derive $g(t_k)$ and $H(t_k)$ respectively as:

$$
g(t_k) = 2 e(t_k) \times \frac{\partial e(\Delta_i)}{\partial \Delta_i} = 2e(t_k)\bar{\tau}_i(t_{k}) \ \text{and;}
$$
$$
H(t_k) = 2 \tau_i(t_k) \times \frac{\partial e(\Delta_i)}{\partial \Delta_i}  = 2\bar{\tau}^{2}_i(t_{k}) 
$$

Since $$ \frac{\partial e(\Delta_i)}{\partial \Delta_i} = \frac{\partial }{\partial \Delta_i}[\Delta_i(t_k^{+})\bar{\tau}_i(t_k) - kB - \beta_k] = \bar{\tau}_i(t_k)$$

Since $H(t_k)$ is in the scalar form we write $[H(t_k)]^{-1} = \frac{1}{2\bar{\tau}^{2}_i(t_{k}) }$

Hence the update equation in \ref{eq9} can be rewritten as:

\begin{equation} \label{eq10}
\Delta_i (t_k^{+}) = \Delta_i (t_k) - \mu \frac{e_i(t_k)}{\bar{\tau}_i(t_k)}
\end{equation}

Since a constant accessing of $\bar{\tau}_i(t_k) = \tau_i(t_k^{+}) - \tau_i (t_k)$ might over-complicate  the synchronization algorithm, we approximate $\bar{\tau}_i(t_k)$ with its mean value, i.e., 

$$
\bar{\tau}_i(t_k) \simeq E[\bar{\tau}_i(t_k)] 
$$

And from the definition of the hardware clock model given by (\ref{eq1}),
$\bar{\tau}_i(t_k) = \int_{t_k^{+}}^{t_{k+1}} f(\zeta)d\zeta$ and hence we can write

\begin{equation}\label{eq13}
\bar{\tau}_i(t_k) \simeq E[\int_{t_k^{+}}^{t_{k+1}} f_i(\zeta)d\zeta] = \int_{t_k^{+}}^{t_{k+1}} E[f_i(\zeta)]d\zeta 
\end{equation}
But,
$$
E[f_i(t)] = \frac{\hat{f} - f_{max} + \hat{f} + f_{max}}{2}  = \hat{f}
$$

\begin{equation}\label{eq14}
\bar{\tau}_i(t_k) \simeq E[\int_{t_k^{+}}^{t_{k+1}} f(\zeta)d\zeta] = \hat{f}\int_{t_k^{+}}^{t_{k+1}} d\zeta  = B\hat{f}
\end{equation}

Therefore the synchronization clock update rule can be simplified as follows:\\

\begin{equation}\label{eq15}
c_i(t_k^{+}) = c_i(t_k) + e_i(t_k) = t_k + \beta_k, \ \text{and}
\end{equation}

and the logical clock rate update as follows:  
\begin{equation}\label{eq16}
\Delta_i (t_k^{+}) = \Delta_i (t_k) - \mu \frac{e_i(t_k)}{B\hat{f}}
\end{equation}

\subsection{Steady State and Convergence Analysis}
Say $e(t_k) = e(k)$ and $\Delta_i(t_k) = \Delta_i(k)$, based on the update equations, (\ref{eq15}) and (\ref{eq16}), we can define $e(k+1)$ and $\Delta_i(k+1)$ as: 

\begin{equation} \label{eq17}
e(k+1) =  \Delta_i(k)\bar{\tau}_i(k) - (B + \beta_{k+1} - \beta_k)
\end{equation}

$$
\Delta_i(k+1) =  \Delta_i(k) - \frac{\mu}{B\hat{f}}[\Delta_i(k)\bar{\tau}_i(k) - (B + \beta_{k+1} - \beta_k)]
$$

\begin{equation} \label{eq18}
\Delta_i(k+1) =  \Delta_i(k)[1 - \frac{\mu}{B\hat{f}}\bar{\tau}_i(k)] - \frac{\mu}{B\hat{f}}(B + \beta_{k+1} - \beta_k)]
\end{equation} 

We can combine (\ref{eq17}) and (\ref{eq18}) into a state representation  given by (\ref{eq19}).

\begin{equation}\label{eq19}
\begin{aligned}
\left[ \begin{array}{c} e(k+1) \\ \Delta_i(k+1) \end{array} \right] =   \begin{bmatrix} 0 & \bar{\tau}_i(k)  \\  0 & 1 - \frac{\mu}{B\hat{f}}\bar{\tau}_i(k)  \end{bmatrix} \left[ \begin{array}{c} e(k) \\ \Delta_i(k)  \end{array} \right] + \\
(B + \beta_{k+1} - \beta_k) \left[ \begin{array}{c} -1 \\ \frac{\mu}{B\hat{f}}  \end{array} \right] 
\end{aligned}
\end{equation} 
%
%

To assess the pairwise convergence of the proposed system in the mean-sense, we evaluate the expectation of \ref{eq19}.

%

%

\begin{equation}\label{eq21}
\left[ \begin{array}{c} E[ e(k+1)] \\ E[\Delta_i(k+1)] \end{array} \right] =   \begin{bmatrix} 0 & B\hat{f}  \\  0 & 1 - \mu  \end{bmatrix} \left[ \begin{array}{c} E[e(k)] \\ E[\Delta_i(k)]  \end{array} \right] +   \left[ \begin{array}{c} -B \\ \frac{\mu}{\hat{f}}  \end{array} \right] 
\end{equation} 

For the algorithm to converge asymptotically in the mean-sense, the modes $\lambda_1,\lambda_2$ of $E[\Phi (k)] = \bar{\Phi} (k)$ must lie in a unit circle \cite{kar_distributed_2009}. Based on the definition $|\lambda I - \bar{\Phi} (k)| = 0$, we obtain the eigenvalues as $[\lambda_1,\lambda_2] = [0,(1-\mu)]$. 

Therefore, a necessary and sufficient condition for asymptotic convergence in the mean-sense is if $0<\mu<2$. Since this condition on convergence is independent on the modes of $\Phi (k)$, a clear advantage over the Gradient Descent Time Synchronization Algorithm \cite{yildirim2016gradient}  and the Proportional Integral Synchronization Algorithm \cite{yildirim_pisync} where the conditions for convergence are respectively, $0<\mu<\frac{1}{B^2 \hat{f}^2}$ and $0<\mu<\frac{2}{B \hat{f}}$ as shown in Table \ref{tab:step_comp}.

\begin{table}[ht!]
	\centering
	\caption{Synchronization Convergence Conditions}
	\begin{tabular}{ll}
		\toprule
		Algorithm  & \multicolumn{1}{l}{Step Size} \\
		\midrule
		GraDeS  \cite{yildirim2016gradient} & $0<\mu<\frac{1}{B^2 \hat{f}^2}$ \\
		PISync \cite{yildirim_pisync} & $0<\mu<\frac{2}{B \hat{f}}$ \\
		Proposed &  $0<\mu<2$\\
		\bottomrule
	\end{tabular}%
	\label{tab:step_comp}%
\end{table}%

The steady state error and clock rate variables can be written as:
$$
\lim_{k\to\infty} E[e(k)] = e(\infty) \ \text{and} \lim_{k\to\infty} E[\Delta_i(k)] = \Delta_i(k)(\infty) 
$$

From equation \ref{eq21}, we can write the respective steady state equations of $e(k) \ \text{and} \ \Delta_i(k)$ as:
$$
\Delta_i(\infty) = (1-\mu)\Delta_i(\infty) + \frac{\mu}{\hat{f}} \Longrightarrow \Delta_i(\infty) = \frac{1}{\hat{f}} \ \text{and}
$$
$$
e(\infty) = B\hat{f}\Delta_i(\infty) - B \Longrightarrow e(\infty) = 0
$$

We therefore conclude from the above analysis that, the synchronization error,$e_i$ of a node $i$ synchronizing to the gateway node $g$,  converges to zero and its logical clock rate,$\Delta_i$ converges to the nominal oscillation period, $\frac{1}{\hat{f}}$.

Let $D_{k+1} = \beta_{k+1} - \beta_k$, $z_k = \Delta_i (k) \hat{f} - 1$ and $w_{k+1} = \int_{t_k}^{t_{k+1}} f(\zeta)d\zeta$
\begin{equation}
\bar{\tau}(k) = B\hat{f} + w_{k+1}
\end{equation}

where $w_{k+1}$ has statistics $E[w_{k+1}] = 0$ and $E[w^{2}(k+1)] = B\hat{f}$.

Based on these definitions we can rewrite the error and clock rate recusion equations respectively as:

$$
e_i(k+1) = z_k \left (B + \frac{w_{k+1}}{\hat{f}} \right) + \frac{w_{k+1}}{\hat{f}} - D_{k+1}
$$
and
$$
\Delta_i(k+1) = \frac{z_{k}+1}{\hat{f}} - \frac{\mu}{B\hat{f}}\left(z_k\left[B + \frac{w_{k+1}}{\hat{f}}\right] + \frac{w_{k+1}}{\hat{f}} - D_{k+1}\right)
$$

Let $g_{k+1} = B\hat{f} + w_{k+1}$ with mean, $E[g_{k+1}] = B\hat{f}$ and second moment, $E[g^{2}_{k+1}] = B^2\hat{f}^{2} + \frac{Bf^{2}_{max}}{3}$

$$
z_(k+1) = z_{k}\left(1 - \frac{\mu}{B\hat{f}}g_{k+1}\right) - \frac{\mu}{B\hat{f}}g_{k+1} + \mu + \frac{\mu}{B}D_{k+1}  
$$
	
\begin{equation}
E[z_(k+1)] = E[z_{k}](1 - \mu)  \Longrightarrow E[z_{\infty}] = 0
\end{equation}
	
$$
\begin{aligned}
z^{2}_{k+1} = z^2_k\left(1 - \frac{\mu}{B\hat{f}}g_{k+1}\right)^2 + \left(- \frac{\mu}{B\hat{f}}g_{k+1} + \mu + \frac{\mu}{B}D_{k+1} \right)^2 + \\
z_{k}\left(1 - \frac{\mu}{B\hat{f}}g_{k+1}\right) \left(- \frac{\mu}{B\hat{f}}g_{k+1} + \mu + \frac{\mu}{B}D_{k+1} \right)
\end{aligned}
$$
							
Taking expectation of both sides, we get
							
\begin{equation}
E[z^2_{k+1}] = E[z^2_{k}]\left( \mu^2 \left[1 + \frac{f^2_{max}}{3B\hat{f}^2}\right] - 2\mu\right) + \frac{\mu f_{max}}{3B\hat{f}^2} + \frac{\mu \sigma^2_D}{B^2}
\end{equation}

Since $E[e_{\infty}] = B E[z_{\infty}] = 0$, it follows that, $Var[e(\infty)] = E[e^2(\infty)] $ and $E[e^2(\infty)]$ can be expressed in terms of $E[z^2(\infty)]$ as:
$$
E[e^2(\infty)] = E[z^2_{k}] \left(B^2 + \frac{E[w^2_{k+1}] }{\hat{f}} \right) + \frac{E[w^2_{k+1}] }{f^2} + + E[D^2_{k+1}]
$$

Using straight-forward steps, the asymptotic variance of the error can be given as:

\begin{equation}
\begin{aligned}
Var[e(\infty)] = \left(B^2 + \frac{Bf^2_{max}}{2\hat{f}}\right)\left(\frac{\mu^2 f_{max} + 2\mu \hat{f}^{2}\sigma^{2}_{D}}{2 - 4\mu - 2\mu^{2} B\hat{f}^2} + 2\mu^{2} f^2_{max}\right)\\
 + \frac{Bf^2_{max}}{f^2} + \sigma^2_D
\end{aligned}
\end{equation}
\section{Time Synchronization Protocol}
In this section, the pairwise synchronization analyses is extended into a fully working implementation protocol for time synchronization in WSNs.
\subsection{Optimal Update Step Size}

We observe that when $\mu = 1$, equation \ref{eq21} becomes

$$
\left[ \begin{array}{c} E[ e(k+1)] \\ E[\Delta_i(k+1)] \end{array} \right] =   \begin{bmatrix} 0 & B\hat{f}  \\  0 & 0  \end{bmatrix} \left[ \begin{array}{c} E[e(k)] \\ E[\Delta_i(k)]  \end{array} \right] +   \left[ \begin{array}{c} -B \\ \frac{1}{\hat{f}}  \end{array} \right] 
$$

where the mean logical clock rate, $E[\Delta_i(k+1)] = \frac{1}{\hat{f}} = \Delta_i(\infty)$. Hence with a unity step size, the steady state clock rate is reached in a single iteration\footnote{This is a known characteristic  of the Newton's Algorithm for adaptive filters \cite{Sayed:2008:AF:1370975}}. This removes the need for a step size adaptation algorithm like in the Gradient Descent Synchronization and PISync algorithms and therefore a reduced processing and memory requirements.

\subsection{Time Synchronization Protocol Based on Newton Adaptive Algorithm}
To achieve network-wide or global network synchronization, the pairwise synchronization analysis is extended into a generalized protocol which each network node uses to achieve global network synchronization. Firstly, since fastest convergence is achieved at unity step size, we adopt this value for the protocol design. Next, since only a few nodes might have access to the gateway node, $g$ and by extension, $t_g$, a distributed averaging process is used to synchronize all nodes to $g$. Finally, the Newton's adaptive search algorithm is used to update the clock rate, $\Delta$, which is used to adjust the speed of the logical clock in each synchronization round. The pseudo-code adopted in the protocol implementation for node $i$ is presented in Algorithm \ref{alg1}.

\begin{algorithm}
	\caption{Synchronization Pseudo-code for Node $i$}\label{alg1}
	\begin{algorithmic}[1]
		\State \textbf{Initialization}
		\State  \ Clear Repository
		\State  \ $c_i \gets 0; ClockError \gets 0; Received \gets 0$;
		\State  Node $i$ broadcasts clock request packets to 1-hop neighbors, $\mathcal{N}_i$
		\State Say node $j$, is such that $\{j\in \mathcal{N}_i| \mathcal{N}_{i} \in \mathcal{V} \ \text{and} \ \mathcal{E}_{i,j} \in \mathcal{E}\}$ 
		\If{<<$c_j$>> request is received at node $j$} 
		\State $j$ sends an acknowledgment with payload <<$c_j$>> 
		\EndIf
		\Comment {<<$c_j$>> denotes packet with payload $c_j$} 
		\For{Each received <<$c_j$>> at node $i$}
		\State $e_{ij} \gets (c_j - c_i)$		\Comment{Nodes $i,j$ neighborhood error}
		\State $ClockError \gets ClockError + e_{ij} $ 
		\State $Received \gets Received + 1$
		\State \textbf{After $T_s$ seconds}		
		\State $e_{new} \gets ClockError / Received$		
		\If{$|e_{new} - c_i|<e_{max}$}
		\State $\Delta_i  = \Delta_i - \mu (e_{new} / \bar{\tau}_i)$
		\EndIf
		\State $c_i\gets c_i + e_{new}$
		\State $ClockError \gets 0; Received \gets 0;$
		\State Set timer to fire after B seconds
		\EndFor
		\State \textbf{Upon receiving <<$c_i$>> clock request from node $l$}\\ \Comment{Given that $i\in \mathcal{N}_l$}		
		\State Node $i$ transmits acknowledgment <<$c_i$>> to node $l$
		\State \textbf{Upon timer, time-out}
		\State Broadcast request for clock packets from $\mathcal{N}_i$ neighbors
		
	\end{algorithmic}
\end{algorithm}

An activated node $i$, with $\mathcal{N}_i$ nearest neighbors, initiates the clock estimate variable, $c_i$, an estimate variable of the oscillator frequency, $\Delta_i$ and two \textit{dummy} variables $ClockError$ and $Recieved$ used for the distributed averaging which are initialized to zero (Line 3). First, node $i$ clears it's initially stored data and broadcast a request for neighborhood time values (Line 4). Now, say node $j$ lies within the neighborhood of $i$ and receives this request, node $j$ replies to $i$ with an acknowledgment packet containing its clock value, $c_j$ (Lines 5-7). For each received acknowledgment with clock value, say $c_j$, node $i$ computes the difference between the received time value, $c_j$ and its current time value, $c_i$. This value is accumulated in the $ClockError$ variable for all neighborhood time value. The $Received$ variable is also incremented by one upon each reception (Lines 8-11). It is assumed that clock values from 1-hop neighbors are received in a short  time, therefore, node $i$ waits a for a certain time, $T_s\lll B$ \footnote{where, $T_s$ is the upper bound on the variance in the convergence time of all network nodes. Within this time, node $i$ is expected to receive values from all nearest neighbors} and computes the mean neighborhood error, $e_{new}$ (Line 13). If $|e_{new}|$ is less than a certain pre-calculated maximum error, say $e_{max}$ \footnote{In \cite{yildirim_pisync}, this value is given by, $e_{max} = 2B\hat{f}^{-1}f_{max}$}, the node updates, $\Delta_i$ using the derived update rule given by \ref{eq10} (Line 15). Node $i$ then updates is clock value, $c_i$, initializes the \textit{dummy} parameters set a timer to fire after B seconds (Lines 16-18). When this timer fires, node $i$ broadcasts request packets for neighborhood clock values. Finally, whenever it received a clock request for say node $l$ localized with its neighborhood, node $i$ transmits an acknowledgment to $l$ with its updated clock.
\section{Practical Experimentation}
In this section, practical experimental evaluation is carried-out for our proposed protocol as against the Gradient Descent Synchronization Protocol \cite{yildirim_gradient_2015}. For each experiment, we evaluate the synchronization accuracy and the convergence time. The experimental setup, methodology and materials are also discussed.
\subsection{Test Network}
\begin{figure}[ht!] 
	\centering    
	\includegraphics[width=0.5\textwidth]{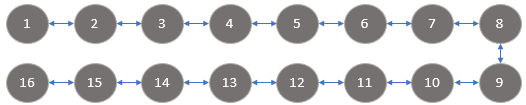}
	\caption[Node Distribution of Sensor Node for Practical Experiments]{Distribution of Sensor Nodes for Experiments}
	\label{fig:linea}
\end{figure}

In our experiments, we utilize a line topology of 16 nodes shown in Figure ~\ref{fig:linea}. Experiments on line topologies allows for the scalability of protocols to be observed since the performance of time synchronization protocols degrade with increase in network diameter ~\cite{energy1}. It should be noted here that, in order to realize these network configurations for our experiments, network nodes are forced to accept packets from only some nodes.

\subsection{Hardware Platform}
In our experiments, the platform based on a network comprising MicaZ nodes from Memsic, instrumented with a 7.37MHz 8-bit Atmel Atmega128L microcontroller are employed. The MicaZ nodes are equipped with 4kB RAM, 128kB program flash and Chipcon CC2420 radio chip which provides a 250 kbps data rate at 2.4 GHz frequency. The 7.37MHz quartz oscillator on the MicaZ board is employed as the clock source for the timer used for timing measurements. This timer operates at 1/8 of that frequency and thus each timer tick occurs at approximately every 921 kHz (approx. 1 $\mu s$). TinyOS-2.1.2 installed on Ubuntu Linux Distribution 14 is used a the base operating system for all experimental work. 

The CC2420 transceiver on the MicaZ board employed has the capability to timestamp synchronization packets at MAC layer with the timer used for timing measurements. Packet level time synchronization interfaces provided by TinyOS are utilized to timestamp synchronization messages at MAC layer ~\cite{Maroti_ttp}.


\subsection{Experiment Parameters}
In our experiments, a beacon period, $B$ of 30 seconds and $e_{max}$ of 6000 $ticks$ were used for both protocols. The nominal oscillator frequency of $\hat{f} =  1MHz$ is used.   Node '1' is used as the reference node for both protocols. The TMicro timer is adopted for all timing operations. When each experiment commences, network nodes are switched on randomly within 5 minutes of operation with each experiment taking about 3.4 hours.

%

\subsection{Results}
In this section, we look at the behavior of the suggested protocol in terms of global synchronization accuracy and convergence time in comparison to the Gradient Descent Synchronization Protocol (GradSync) and Average Proportional Integral  Synchronization Protocol (AvgPISync). These metrics of evaluation are thoroughly defined by Yildirim \textit{et \ al} \cite{yildirim_pisync}.

\begin{table}[h!]
	\centering
	\caption{Convergence Time}
	\begin{tabular}{lrrr}
		\toprule
		\multicolumn{1}{c}{\textbf{Protocols}} & \multicolumn{1}{c}{GraDeS} & \multicolumn{1}{c}{AvgPISync} & \multicolumn{1}{c}{Proposed} \\
		\textbf{Conv. Time($s$)} & \multicolumn{1}{c}{545} &\multicolumn{1}{c}{660} & \multicolumn{1}{r}{215} \\
		\bottomrule
	\end{tabular}%
	\label{tab:compLine}%
\end{table}%

\begin{figure}[ht!] 
	\centering    
	\includegraphics[width=0.5\textwidth]{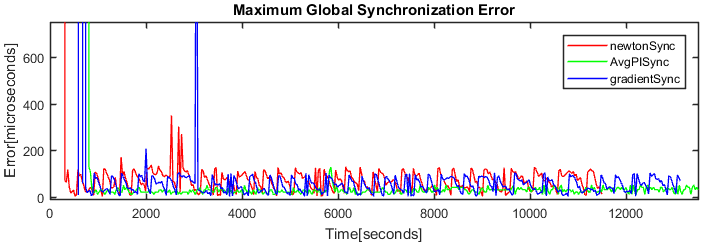}
	\caption[Maximum Global Synchronization Error Comparison]{Maximum Global Error Comparison}
	\label{fig:max_global}
\end{figure}

The maximum global error is presented for FTSP, FloodPISync and BAF in Figure ~\ref{fig:max_global}. Table ~\ref{tab:compLine} summarizes these the convergence time results. 

First, we observe that the proposed protocol outperform both GraDeS and AvgPISync protocols in terms of convergence time. This tallies with the theoretical analyses presented in section II where the clock rate multiplier at unity step size converges at a faster rate as compared to GraDeS. GraDeS and AvgPISync however have a similar performance in terms of the convergence time. Next we observe that, the suggested method performs similarly to AvgPISync but is outperformed by GraDeS in terms of synchronization error. However after achieving convergence, GraDeS experiences an error peak of about $e^{G}_{max} = 1157 \mu s$ at about 3180 seconds. GraDeS protocol outperforms the proposed protocol in terms of synchronization error since its narrow step size bound allows for more tighter synchronization. PISync also with narrow step size than NewtonSync protocol experiences a better synchronization error.

%
%

\section{Conclusion}
In this work, we present a method of time synchronization for wireless sensor networks inspired by the Newton adaptive algorithm. The convergence of the proposed algorithms in the mean sense was derived and the asymptotic mean and variance of pairwise synchronization were derived. A practical protocol is designed by extending the pairwise synchronization analysis. The protocol is compared experimentally against AvgPISync and GraDeS. The proposed scheme outperformed the other protocols in terms of convergence time but performed closely to AvgPISync. Future work could include further testing the protocol on mobile nodes and on larger networks.

\ifCLASSOPTIONcaptionsoff
  \newpage
\fi

\bibliographystyle{ieeetr}
\bibliography{IEEEabrv,ref}

\end{document}